\documentclass[aps,prl,twocolumn,showpacs,superscriptaddress]{revtex4}
\usepackage{amsfonts,amsthm,amssymb}
\usepackage{amsfonts}
\usepackage{amsmath}
\usepackage{amssymb}
\usepackage{graphicx}
\usepackage{dcolumn}
\usepackage{enumitem}
\usepackage[colorlinks,urlcolor=blue,citecolor=blue]{hyperref}

\begin{document}

\title{Twisted kinks, Dirac transparent systems, and Darboux transformations}

\author{
{\bf F. Correa${}^{a}$ and V. Jakubsk\'y${}^{b}$
}
\\
[4pt]
{\small \textit{${}^{a}$ Centro de Estudios Cient\'{\i}ficos (CECs), Casilla 1469, Valdivia, Chile}}\\
{\small \textit{{${}^{b}$}
Department of Theoretical Physics,
Nuclear Physics Institute,
25068 Re\v z, Czech Republic }}\\
\sl{\small{E-mails: correa@cecs.cl, jakub@ujf.cas.cz
} }}


\begin{abstract}

Darboux transformations are employed in construction and analysis of Dirac Hamiltonians with pseudoscalar potentials. By this method, we build a four parameter class of reflectionless systems. Their potentials correspond to composition of complex kinks, also known as twisted kinks, that play an important role in the $1+1$ Gross-Neveu and Nambu-Jona-Lasinio field theories. The twisted kinks turn out to be multi-solitonic solutions of the integrable AKNS hierarchy. Consequently, all the spectral properties of the Dirac reflectionless systems are reflected in a non-trivial conserved quantity, which can be expressed in a simple way in terms of Darboux transformations.
 We show that the four parameter pseudoscalar systems reduce to well-known models for specific choices of the parameters. An associated class of transparent non-relativistic models described by matrix Schr\"odinger Hamiltonian is  studied and the rich algebraic structure of their integrals of motion is discussed.

\end{abstract}

\maketitle

\section{Introduction}

Recently, an important contribution in the theory of exactly solvable one dimensional Dirac Hamiltonians was done by Dunne and Thies \cite{DT}. They found a generic class of transparent, time-dependent Dirac systems whose dynamical equations can be solved algebraically.
A transparent system, also called reflectionless, is peculiar by the absence of any reflection in the scattering modes. The results of \cite{DT} were originally motivated by their connection with Hartree-Fock scattering solutions of $1+1$ dimensional quantum field theories of massless, self-interacting Dirac fermions \cite{DT2, DT3}. Specifically, in the two dimensional Gross-Neveu (GN$_2$) \cite{GN} and Nambu-Jona-Lasinio (NJL$_{2}$) models \cite{NJL},  the self-consistent mean field potentials turn out to be transparent in the case of scattering, static or time-dependent, solutions \cite{DT2, DT3, TN1, TN2}. This means that the GN$_2$ and NJL$_{2}$ models can be solved analytically by using the Hartree-Fock approximation, where the dynamics is reduced into a single particle Dirac equation, accompanied by an additional requirement of self-consistency, see for example \cite{DT2,TN1,dashen}. In this context, the transparent scalar and pseudo-scalar potentials in Dirac equation can be treated as fermion 
condensates and correspond to scattering solutions of the gap 
equation in the GN$_2$ and NJL$_{2}$ models \cite{DT,DT2, DT3,TN1}. 

Dunne and Thies generalized the result of Kay and Moses \cite{KM} on stationary Schr\"odinger operators with transparent potentials, obtained by pure algebraic methods. A similar result was also provided by Takahashi \emph{et. al.} in refs. \cite{TN1,TN2,TN3}, but for a generic class of \emph{static} transparent Dirac potentials.  In this way, static scattering Hartree-Fock solutions of GN$_2$ and NJL$_{2}$ were found. Those time-independent  solutions are closely related with the modified Korteweg-de Vries (mKdV) and Ablowitz-Kaup-Newell-Segur (AKNS) hierarchy of integrable systems, respectively \cite{BDT, CDP}. For a more detailed discussion of the advances in this topic we refer to \cite{DT, DT3} and references therein. 

In the present work, we will consider the construction of static reflectionless systems that is alternative to the approach utilized in refs. \cite{DT,DT2, DT3,TN1, TN3}.
We will describe how  transparent (reflectionless) time-independent potentials, and therefore self-consistent condensates in the language of GN$_2$ and NLJ$_{2}$,  can be obtained by means of Darboux transformations \cite{DiracDarboux, susytwisting}. As an illustration we will build up a 2-twisted kink,   also called 2-soliton, pseudoscalar potential and examine its properties. Additionally, we will show how the well-known Hartree-Fock static solutions of the GN$_2$ and NJL$_{2}$, e.g. the real  Coleman-Callan-Gross-Zee kink \cite{dashen}, the Shei's twisted kink \cite{shei} and the real Dashen-Hasslacher-Neveu baryon \cite{dashen} can be directly derived from the 2-twisted kink and the Darboux transformations.

The Darboux transformations represent a powerful tool in quantum physics. It allows to construct a Hamiltonian together with the solutions of the associated stationary equation from the eigenstates of an initial, known,  Hamiltonian. The interactions described by the new and old Hamiltonians can be completely different, it depends on the choice of the Darboux transformation. This freedom allow us to obtain new and interesting models from a known one. In the non-relativistic regime of a particle governed by the Schr\"odinger equation, the Darboux transformation forms the back-bone of the standard supersymmetric quantum mechanics \cite{susyreferences}. Darboux transformations provide a natural explanation of the intrinsic properties of reflectionless systems by their relation with the free particle system: all the static transparent potentials can be generated by a chain of Darboux transformations  from the free particle Schr\"odinger Hamiltonian  \cite{Matveev}.

Here, we utilize Darboux transformations in a similar manner for the construction of (pseudo-)scalar transparent potentials in Dirac Hamiltonians. As a consequence of the realization of such  
Darboux transformations, the attibutes of reflectionless Dirac Hamiltonians are revealed. A higher order integral of motion in quantum systems with transparent potentials can be obtained by dressing of the translation symmetry operator of the free-particle system. This conserved quantity coincides with the higher order matrix differential Lax operator of the AKNS hierarchy.  Together with the Hamiltonian, they compose the so called Lax pair \cite{Gesz} of a corresponding hierarchy of stationary non-linear integrable equations. Non-relativistic reflectionless systems are associated with the Korteweg-de Vries (KdV) hierarchy whereas the Dirac Hamiltonians with scalar reflectionless potentials with the mKdV hierarchy.  The general family of static transparent $(1+1)$ dimensional Dirac systems associated with the mKdV hierarchy was analyzed recently in  \cite{MikhailAdrian} in the context of nonlinear 
supersymmetry with the use of Darboux transformation.  Here, we will consider Dirac Hamiltonians with pseudo-scalar potentials related with the AKNS hierarchy.  The presence and features of a non-trivial conserved quantity will be also discussed. Such integral of motion is fundamental in transparent static systems;  it reflects spectral properties, as for instance the degeneracy of energy levels.

The paper is organized as follows: in Section \ref{sec2} we briefly review the framework of Darboux transformation for one-dimensional Dirac Hamiltonians. We discuss a class of solvable models in Section \ref{sec3} where the four parameter twisted kink potentials are constructed.  Their link with the AKNS hierarchy of nonlinear equations is explained.  Moreover, we show that for the specific values of parameters, they coincide with well-known solutions  in the context of the GN$_2$ and NJL$_2$ models. In section \ref{sec4}, we consider non-relativistic systems described by Schr\"odinger Hamiltonian with matrix potential.  We show that a rich algebraic structure closed by integrals of motion accompanies these non-relativistic systems with transparent matrix potentials. The last section is devoted to discussion and outlook.

\section{Darboux transformations in Dirac Hamiltonians}\label{sec2}

Let us consider the one-dimensional Dirac Hamiltonian
\begin{equation}\label{h}
H=i\sigma_2 \partial_x+\Sigma(x) \, \sigma_1+M(x) \sigma_3 \, ,
\end{equation}
with the scalar $\Sigma(x)$, and pseudo-scalar $M(x)$ potentials \cite{footnote1}. 
The dynamics of the quantum system is governed by the stationary equation 
\begin{equation}\label{stac}
 H\Psi=E \Psi\, .
\end{equation}
This equation appears as an effective dynamical equation in diverse areas of physics. Besides the above mentioned $1+1$ quantum field theories, it also appears in description of different condensed matter systems. In the context of superconductivity, the Hamiltonian in (\ref{stac}) is named  Bogolioubov -de Gennes (BdG) Hamiltonian and is frequently used in a modified form 
\begin{align}\label{bddg}
H_{\text{BdG}}=\left(\begin{array}{cc}
 i\partial_x & \Delta(x) \\
\Delta^*(x) &  -i\partial_x  \\
\end{array}\right)\,,\quad \Delta(x)=\Sigma+iM \, .
\end{align}
The Hamiltonian $H_{\text{BdG}}$ is unitary equivalent to (\ref{h}) by means of the transformation $H_{\text{BdG}}=\exp \left(i\frac{\sigma_1 \pi}{4} \right)  H \exp \left(-i\frac{\sigma_1 \pi}{4} \right)$. 
In the current paper, we are interested in the application of Darboux transformations in the context of GN$_2$ and NJL$_2$ models, where the eq. (\ref{stac}) and the explicit form of the potentials $M$ and $\Sigma$, and particularly $\Delta$, are relevant for their Hartree-Fock solutions.

Number of solvable systems described by (\ref{h}) is rather limited. Here, we discuss briefly the construction based on the Darboux transformations that allows us to find a new solvable Hamiltonian $H$ of the type (\ref{h}) starting from  an operator $H_0$, where the solutions of $H_0\Psi_0=E\Psi_0$ are known. By construction, the spectra of $H$ and $H_0$ are almost identical despite the fact that the two Hamiltonians can have completely different potentials. We recommend \cite{DiracDarboux} and \cite{susytwisting} for more details.

The Darboux transformation $L$ is a matrix differential operator defined in terms of two eigenstates $u_1=(u_{11},u_{21})^T$ and  $u_2=(u_{12},u_{22})^T$ of $H_0$, where $T$ means transposition. In principle, the eigenstates do not need to be physical (they can be just formal solutions of (\ref{stac})). First we assemble the $2\times 2$ matrix 
\begin{align}
U=(u_1,u_2)=\left(\begin{array}{cc}
u_{11} & u_{12} \\
u_{21} &  u_{22} \\
\end{array}\right),\quad (H_0-E_{1(2)})u_{1(2)}=0 \, .
\end{align}
By construction, the matrix $U$ satisfies
\begin{equation}
H_0U= U\mathcal{E},\quad \mathcal{E}=\left(\begin{array}{cc}E_1&0\\0&E_2\end{array}\right).
\end{equation}
Now, we define the Darboux transformation in the following manner \cite{footnote2}
\begin{equation}\label{L}
 L=U\frac{\partial}{\partial x} U^{-1}=\mathbf{1}\partial_x-U' U^{-1},\quad L\, U=0\, ,
\end{equation}
where the latter equation means that  the operator $L$ annihilates the states $u_1$ and $u_2$, $L \, u_{1(2)} =0$. The operator (\ref{L}) mediates an intertwining relation between $H_0$ and a new Hamiltonian $H$
\begin{equation}\label{intertwining}
 L H_0=H L \,,
\end{equation}
where $H$ can be computed from any of the equivalent formulas
\begin{align}\label{htilde}
 H&=H_0+i\, [\, \sigma_2\, ,\, U'\,U^{-1}\, ]\\&=\sigma_2\, H_0\, \sigma_2+\sigma_2\, [\,\sigma_2\, ,\, U \, \mathcal{E}\, U^{-1}\, ]\label{h2}\\
 &=\sigma_2 \, H_0\, \sigma_2+\left(\frac{u_1^T\sigma_1u_2}{\det U}\sigma_3-\frac{u_1^T\sigma_3u_2}{\det U}\sigma_1\right)(E_1-E_2) \label{h3}\, .
\end{align}
The conjugate operator $L^{\dagger}$ can be written as  $L^{\dagger}=-\partial_x+V'V^{-1}$ 
where $V=(U^{\dagger})^{-1}=(v_1,v_2)$, and satisfies an inverse intertwining relation in comparison to (\ref{intertwining}),
\begin{align}\label{intertwining2}
\quad L^{\dagger} H= H_0 L^{\dagger}\, .
\end{align}
The columns $v_1$ and $v_2$ form the kernel of the operator $L^{\dagger}$, $L^{\dagger}v_{1(2)}=0$, and solve the equation $ H v_{1(2)}=E_{1(2)} v_{1(2)}$. The intertwining operators satisfy the following remarkable identity
\begin{subequations}
\label{poly}
\begin{align}
 L^{\dagger}L&=P(H_0)=(H_0-E_1)(H_0-E_2), \\
LL^{\dagger}&=P(H)=(H-E_1)(H-E_2) \, .
\end{align}
\end{subequations}
These relations differ from those appearing in the standard supersymmetric quantum mechanics. In the latter framework, the anticonmutator of two conjugate supercharges is \textit{linear} in Hamiltonian \cite{susyreferences}, whereas in \eqref{poly}, it is a second order polynomial of $H$. Let us notice that the factorization \eqref{poly} of the intertwining operators, nonlinear in the Hamiltonian, is a landmark of the nonlinear supersymmetry (See for instance \cite{susynonlinrefs} and references therein). By looking at \eqref{poly} it is straightforward to check that both Darboux operators, $L$ and $L^\dagger$ annihilate states of energies $E_1$ and $E_2$.

The interaction described by $H$ is strongly dependent on the choice of $u_1$ and $u_2$. On one hand, $H$ can be unitarily equivalent to $H_0$ as long as we choose $u_{1(2)}$ such that $E_1=E_2$ (then we have $ \mathcal{E} \sim \mathbf{1}$, see (\ref{h2})  or  (\ref{h3})). On the other hand, if we fix $u_1$ and $u_2$ such that $\det U$ ceases to be nodeless, $H$ has additional singularities when compared to $H_0$. We disregard both cases, the first for being trivial while the second one for being non-physical; it produces singularities in the real line \cite{footnote3}. 

The spectrum of the new Hamiltonian can be found with the use of the operator $L$. The eigenstates $\Psi$ of $H$ corresponding to the energy $E$ are given by \cite{footnote4}
\begin{equation}\label{eigen}
 \Psi=L\Psi_0,\quad H\Psi=E\Psi \, ,
\end{equation}
as long as we have $(H_0-E)\Psi_0=0$ and $\Psi_0\neq u_{1(2)}$. 
Hence, the spectra of $H_0$ and $H$ are identical up to possible difference in the eigenvalues $E_1$ and $E_2$. Considering the eigenvalues $E_{1(2)}$, any of the possibilities is possible: both, one, or none of $E_{1(2)}$ are in the spectrum $H_0$ and both, one or none of $E_{1(2)}$ are in the spectrum of $H$. We will illustrate the situation later on where $H$ possesses one or two additional energy levels when compared to $H_0$ (i.e.  $E_{1(2)}$ are missing in the spectrum of $H_0$ but they are in the spectrum of $H$). If $H$ has additional bound states when compared to $H_0$, they correspond to the vectors $v_1$ and $v_2$ from the columns of the matrix
\begin{align}\label{ecv}
V=(U^{\dagger})^{-1}
,
 \quad HV=V\mathcal{E} \, .
\end{align}

\section{2-twisted kinks via Darboux transformations}\label{sec3}

We can utilize the procedure described in the preceding section for the construction of a Hamiltonian $H$ with novel pseudo-scalar potentials. The initial  Hamiltonian $H_0$ depends only on a constant mass parameter $m$ and reads
\begin{equation}\label{free}
H_0=i\sigma_2 \partial_x+m\, \sigma_3 \, .
\end{equation}
This Hamiltonian  represents a free system. Its spectrum consists of two separated  energy bands $E\in (-\infty,-m] \cup [m,\infty)$. The associated scattering wavefunctions,
\begin{subequations}
\label{freepsi}
\begin{align}
\psi_{\pm k}^{\rightarrow}& =\left(
\begin{array}{c}
  \frac{ik}{m\mp \sqrt{k^2+m^2}}\\
 -1 \\
\end{array}
\right) e^{ikx}, \\
\psi_{\pm k}^{\leftarrow} &=\left(
\begin{array}{c}
  \frac{ik}{m\mp \sqrt{k^2+m^2}}\\
 1 \\
\end{array}
\right) e^{-ikx} \, .
\end{align}
\end{subequations}
correspond to the eigenvalues $E=\pm \sqrt{k^2+m^2}$ for positive and negative $k$ index respectively.
The model described by (\ref{free}) is reflectionless; the wave functions \eqref{freepsi} do not contain any backscattered components for all energy values.

\begin{figure}
\includegraphics[scale=.8]{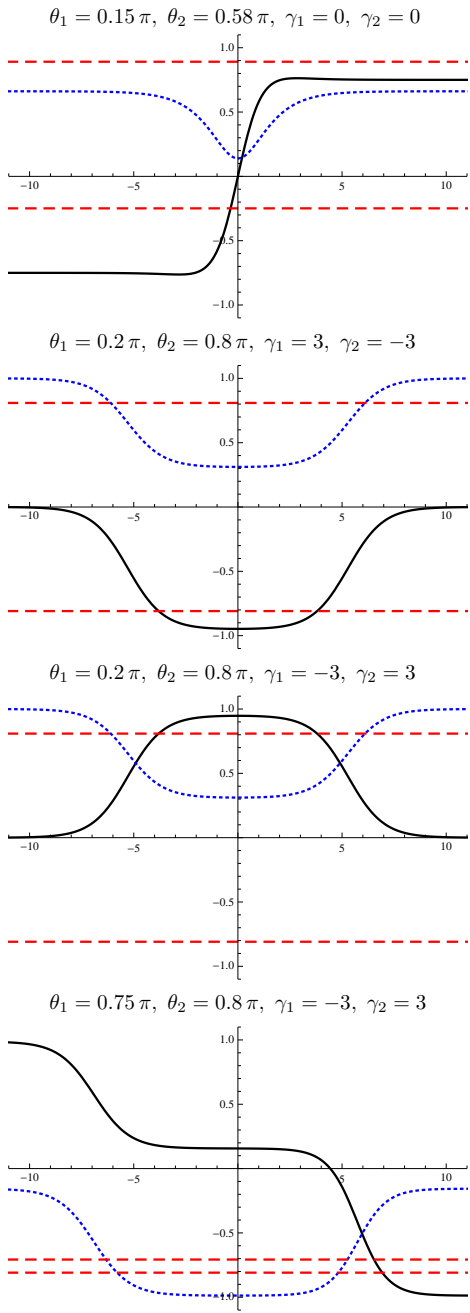}
\caption{The solid (black) line corresponds to $\Sigma$ and dotted (blue) line corresponds to $M$. Dashes (red) lines correspond to the energies $E_1$ and $E_2$. The threshold of the continuum spectrum is at $\pm m$, where we fixed $m=1$. The selected values of the parameters are denoted in each plot.}\label{fig1}
\end{figure}

In order to define a Darboux transformation 
we fix the seed vectors $u_1$ and $u_2$ as
\begin{subequations}
\label{b11}
\begin{align}
u_1&=\left(
\begin{array}{c}
  -\cos \frac{\theta_1}{2}  \cosh \left(m\, \sin\theta_1 \, x +\gamma_1\right) \\
 \sin \frac{\theta_1}{2} \sinh \left(m\,    \sin\theta_1\, x +\gamma_1\right) \\
\end{array}
\right), \\ u_2&=\left(
\begin{array}{c}
  -\cos \frac{\theta_2}{2} \sinh \left(m\,  \sin\theta_2 \, x +\gamma_2\right) \\
\sin \frac{\theta_2}{2}   \cosh \left(m\,   \sin\theta_2 \, x +\gamma_2\right)\\
\end{array}
\right),
\end{align}
\end{subequations}
where the free parameters are taken from the range $\gamma_1, \gamma_2 \in(-\infty,\infty)$ and $\theta_1, \theta_2 \in[0,\pi]$. Despite the spinors $u_1$ and $u_2$ satisfy
\begin{align}
H_0 \, u_1 =m\,  \cos \theta_1 \, u_{1} , \quad H_0 \, u_2 =m\,  \cos \theta_2 \, u_2 \, ,
\end{align} 
they do not represent physical states as they diverge for large $x$. As long as the seed vectors $u_1$ and $u_2$ are fixed, the new Hamiltonian $H$  can be found directly, using any of the relations in (\ref{htilde}). As a result, $H$ has the following effective vector potential $\Sigma$ and the effective-mass term $M$,
\begin{widetext}
\begin{align}\label{sigma}
\Sigma&=m\left(\cos \theta_2-  \cos \theta_1 \right)\frac{\displaystyle \tan \frac{\theta_1}{2} \tanh\left( m\, \sin\theta_1 \,  x +\gamma_1 \right)- \cot \frac{\theta_2}{2}  \tanh\left( m\, \sin\theta_2  \, x +\gamma_2 \right)}{\displaystyle1-\tan \frac{\theta_1}{2} \cot \frac{\theta_2}{2} \tanh\left( m\, \sin\theta_1\,  x +\gamma_1 \right)\tanh\left(m\, \sin\theta_2  \, x +\gamma_2 \right)}, \\
\label{eme}
\frac{M}{m}&=-1+\cos \theta_2-  \cos \theta_1 -\frac{\displaystyle 2\left( \cos \theta_2-  \cos \theta_1  \right) }{\displaystyle 1-\tan \frac{\theta_1}{2} \cot \frac{\theta_2}{2} \tanh\left( m\, \sin\theta_1\,  x +\gamma_1 \right)\tanh\left(m\, \sin\theta_2  \, x +\gamma_2 \right) }.
\end{align}
\end{widetext}
The Hamiltonian (\ref{h}) with (\ref{sigma}) and (\ref{eme}) is physically relevant when the pseudo-potential does not have any singularities in the real line, i.e. the determinant of the matrix $U$ is nodeless, see (\ref{htilde}). Its explicit form
\begin{align}\label{detU}
&\det U= \\
&\frac{1}{2} \sin \left(\frac{\theta_1-\theta_2 }{2}\right) \cosh (\gamma_1+\gamma_2+m x\, ( \sin \theta_1+ \sin\theta_2))+\notag \\
&-\frac{1}{2} \sin \left(\frac{\theta_1+\theta_2 }{2}\right)  \cosh (\gamma_1-\gamma_2+m x\, ( \sin \theta_1-\sin\theta_2) ), \notag 
\end{align}
suggests that, to rule out the singular potentials, it is sufficient to fix $\theta_1$ and $\theta_2$ in the following manner
\begin{align}\label{reg}
0\leq\theta_1 < \theta_2\leq\pi\, ,
\end{align}
which we will suppose to be the case from now on. The case $0<\theta_1=\theta_2<\pi$ is not interesting as the determinant (\ref{detU}) is constant and $H$ is equivalent to $H_0$. In the Figure \ref{fig1}  we present in the table the potential terms (\ref{eme}) and (\ref{sigma}) for some explicit choices of the parameters. Later on, some specific models will be discussed in more detail.

In contrast with the free particle case, the Hamiltonian $H$ can have up to two bound states. They are represented by $v_1$ and $v_2$, see (\ref{ecv}), 
\begin{subequations}
\label{v} 
\begin{align}
v_{1}&=\frac{1}{\det U}\left(\begin{array}{r} \sin \frac{\theta_2}{2}\cosh(m \sin \theta_2\, x+\gamma_2)\\ \cos \frac{\theta_2}{2}\sinh(m \sin \theta_2\,  x+\gamma_2)\end{array}\right), \\ 
v_{2}&= \frac{1}{\det U}\left(\begin{array}{r} \sin \frac{\theta_1}{2}\sinh(m \sin \theta_1\, x+\gamma_1)\\ \cos \frac{\theta_1}{2}\cosh(m \sin \theta_1\,  x+\gamma_1)\end{array}\right).
\end{align}
\end{subequations}
and satisfy \cite{footnote5} $H v_a=E_a v_a$ with $E_1=m\cos \theta_1$ and $E_2=m\cos \theta_2$.

Both spinors $v_1$ and $v_2$ are square integrable for $\theta_1,\theta_2\in(0,\pi)$  and therefore $H$ has two bound states with energies $E_1$ and $E_2$. For $\theta_1=0$ and $\theta_2<\pi$, $v_1$ ceases to be square integrable while $v_2$ is a bound state. In this case, $E_1$ belongs to the bottom of the positive continuum spectrum. A similar situation occurs when $\theta_1>0$ and $\theta_2=\pi$. Here, $v_2$ is not square integrable while $v_1$ is a bound state. $E_2$ forms the threshold of the negative energies. In both cases, $H$ will have one bound state with energy $E_1=m\cos\theta_1$ or $E_2=m\cos\theta_2$. When $\theta_1=0$ and $\theta_2=\pi$, neither $v_1$ nor $v_2$ are bound states and $H$ is isospectral with $H_0$.

The scattering sector of the system can be obtained by using the intertwining operator $L$ in eq. (\ref{eigen}) on the functions defined in \eqref{freepsi}, yielding
\begin{align}
\Psi_{\pm k}^{\rightarrow}=L\psi_{\pm k}^{\rightarrow}\quad , \quad \Psi_{\pm k}^{\leftarrow}=L\psi_{\pm k}^{\leftarrow}\,.
\end{align}
The pseudo-scalar potential of $H$ is reflectionless for all energy values. Since $U'U^{-1}$ converges to a constant matrix for large $x$,  the Darboux transformation applied on the incoming or outgoing states of the free system \eqref{freepsi} generates incoming or outgoing state of the system described by $H$.

Let us discuss additional consequences of the transparent nature of the pseudo-scalar potentials. For these purposes is  convenient to use the function $\Delta(x)=\Sigma+i M$ (\ref{bddg}) which usually appears in the analysis of the GN$_2$ and NJL$_2$ systems in BdG Hamiltonians, see refs. \cite{DT2, DT3, CDP}. The analysis of the asymptotic behavior of $\Delta$ at large $|x| \rightarrow \infty$ gives the following formula
\begin{align}\label{asymptotic}
\Delta(x=+\infty)=e^{2i (\theta_1 +\theta_2 )} \Delta(x=-\infty)\, ,
\end{align}
which allows to call $\Delta$ as a 2-twisted kink \cite{DT3}. The net rotation of the 2-twisted kink through the real line is just an angle $2(\theta_1 +\theta_2 )$. The formula reflects existence of bound states by a kink or an antikink profile of $\Delta$, in dependence on the actual values of $\theta_1$ and $\theta_2$. 
The separation between the twisted (anti)kinks is related with the parameters $\gamma_1$, $\gamma_2$, $\theta_1$ and $\theta_2$ see Figure \ref{fig1}. 
The formula (\ref{asymptotic}) suggests that twisted kinks  for $n>2$ can be obtained by a chain of consecutive Darboux transformations applied to the free particle.  The situation should be analogous to the multi-soliton solutions in the Schr\"odinger case, related with the KdV hierarchy \cite{Matveev}.

The peculiar properties of $\Delta(x)$ follow from the fact that satisfies a complex equation of the AKNS hierarchy
\cite{Gesz}
\begin{widetext}
\begin{equation}\label{akns}
\Delta'''-6|\Delta|^2\Delta'+2 im \left(\cos \theta_1+\cos \theta_2 \right)(\Delta''-2|\Delta|^2\Delta)+2m^2(1-2 \, \cos \theta_1\cos \theta_2)\Delta'+4 im^3 \left(\cos \theta_1+\cos \theta_2\right)\Delta=0\,.
\end{equation}
\end{widetext}

The function $\Delta^*$ solves the equation conjugated to (\ref{akns}). Hence,  $\Delta$  is a 2-soliton solution, or a 2-twisted kink. It is worth remarking that this specific multi-soliton was discussed in \cite{TN3} and was also generalized to the time-dependent case in \cite{DT3}. 
Eq. (\ref{akns}) also provides another verification that the potential is reflectionless; it is known that potentials of scattering nature satisfying one of the equations of the AKNS hierarchy have this property \cite{Gesz}. For instance, the free particle potential $\Delta_0(x)=im$ satisfies the equation $\Delta_0'=0 $ that correspond to the first equation  in the AKNS hierarchy.
In the context of GN$_2$ and NJL$_2$ models, static Hartree-Fock multi-solitons of scattering nature satisfy an equation of the mKdV and AKNS hierarchy respectively. The same argument also applies for the case of periodic solitons as the crystal kink and its complex version \cite{BD}. Moreover, concerning time dependent solutions of the GN$_2$ and NJL$_2$ models, the 2-twisted kink can be trivially boosted with a velocity $v$ as $\Delta(x) \rightarrow \Delta(x-vt)$. 

A Dirac Hamiltonian, whose pseudoscalar potential $\Delta$ satisfies one of the equations of the AKNS hierarchy, has a non-trivial integral of motion $Q$ that can be identified with the Lax operator. In this sense,  $Q$ and $H$ compose the Lax pair of the AKNS hierarchy \cite{Gesz}. A standard definition of $Q$ is given in terms of the functions $f_n$ and $g_n$
\begin{align}\label{gh}
f_n=\sum_{k=0}^n c_{n-k} \hat{f}_k,  \quad g_n=\sum_{k=0}^n c_{n-k} \hat{g}_k \, ,
\end{align}
where $\hat{f}_k$ and $\hat{g}_k$ satisfy the recursion relations
\begin{align}
\hat{f}_n=\frac{i}{2} \hat{f}'_{n-1}+\Delta\, \hat{g}_n, \quad  \hat{g}'_n=-i \Delta^*  \hat{f}_{n-1}+ i \Delta  \hat{f}_{n-1}^* \, .
\end{align}
Here, $\hat{g}_0=1$ and $\hat{f}_{-1}=0$ and $c_m$ are real constants, $c_0=1$. Thus, $f_n$ and $g_n$ are functions of $\Delta$ and of its derivatives. The higher-order, nonlinear differential equations for $\Delta$ and $\Delta^*$, resulting from $f_n=f^*_n=0$, form the AKNS hierarchy, for more details see \cite{Gesz}. In case of our 2-twisted kink solution $\Delta$ with $M$ and $\Sigma$ given by (\ref{eme}) and (\ref{sigma}), the l.h.s of (\ref{akns}) coincides with  the function $f_3$ with the constants $c_k$ fixed as $c_1= -\cos \theta_1-\cos \theta_2$, $c_2=  \cos  \theta_1 \cos  \theta_2-\frac{1}{2}$  and $c_3= \frac{1}{2} (\cos \theta_1+\cos \theta_2)$.  It is worth mentioning that as long the function $\Delta$  solves one of the equations of the hierarchy (i.e. $f_n=0$), it solves all the equations of higher order as well ($f_m=0$, $m>n$) for a specific choice of the constants.

Once the functions (\ref{gh}) are fixed and $f_n=0$ for an integer $n$, the Lax operator is given by

\begin{widetext}
\begin{equation}\label{q}
Q=i\sum_{m=0}^n \left( 
\begin{array}{cc}
-\frac{i}{2}\left( f_{n-m-1} +f^*_{n-m-1} \right) & \frac{1}{2}\left( f_{n-m-1} -f^*_{n-m-1} -2i g_{n-m} \right)  \\
 \frac{1}{2}\left( f_{n-m-1} -f^*_{n-m-1} +2i g_{n-m} \right) & \frac{i}{2}\left( f_{n-m-1} +f^*_{n-m-1} \right)  \\
\end{array}  \right) H^m \, ,
\end{equation}
\end{widetext}

and satisfies
\begin{equation}
 [Q,H]=0 \, ,
\end{equation}

 Darboux transformation paves the way for an alternative construction of the Lax operator (\ref{q}) and provides insight into its physical nature. In the free particle system $\Delta=const$, the Lax operator coincides with the momentum operator $p=-i\partial_x\mathbf{1}$,
\begin{equation}
[H_0,p]=0 \, .
\end{equation}
The Lax operator associated with $H$ can be easily constructed directly by ``dressing'' of the momentum $p$ by the Darboux transformations, 
\begin{equation}\label{qh}
Q=Lp L^{\dagger} \, .
\end{equation}
Moreover, it is straightforward to compute the square of the conserved quantity $Q$, 
\begin{align}
Q^2&=Lp L^{\dagger} L p L^{\dagger}
=L (H_0^2-m^2)P(H_0) L^{\dagger} \notag  \\
&=(H^2-m^2) (H-E_1)^2 (H-E_2)^2 \, . \label{q2}
\end{align}
Here we use $H_0^2=p^2+m^2$ and that in the case of a generic polynomial, $ \mathcal{P}(H_0)$, the intertwining relations work identically as in eq. (\ref{intertwining}),  $L \mathcal{P}(H_0)=\mathcal{P}(H)L $. The polynomial $Q^2$ is known as the spectral polynomial \cite{Gesz}. It follows from (\ref{q2}) that $Q$ annihilates the bound states $v_1$ and $v_2$ as well as the scattering states corresponding to $E=\pm m$. These energies are non-degenerated in constrast to the energies $E\in]-\infty,-m) \cup (m,\infty[$ that have double degeneracy. Besides, the spectral polynomial reflects the difference between the energies $E_1$ and $E_2$ and those with the energies $E=\pm m$; the former ones are double roots of (\ref{q2}) while the latter ones are single roots of the spectral polynomial.

It is worth noticing that the dressing method and its relation with hidden (super)symmetries was explored in different scenarios, as in one dimensional (non)relativistic systems \cite{MikhailAdrian,AdS2,varias} as well as in the case of the many-particle Calogero model, see  \cite{qcal} and references therein. There, relations of the kind of Eq. (\ref{q2})  appear in the context of a hidden non-linear bosonized supersymmetry where $Q$ plays the role of a supercharge.

The 2-twisted kink $\Delta=\Sigma+i M$ is reduced to well-known solutions for the GN$_2$ and NJL$_2$ models for certain limits of the parameters: 

\begin{itemize}[itemindent=10pt,leftmargin=0pt]
\item\textit{Shei complex kink}:  Fixing $\gamma_1=\gamma_2=\gamma$ with  either $\theta_1=0,  \theta_2=\theta$ or $\theta_1=\theta, \theta_2=\pi$, we get a single twisted (anti)kink
\begin{align}
\Sigma= \pm m\sin \theta  \tanh \left(m\,   \sin \theta\, x+\gamma\right),\quad M=\mp m\cos \theta\,, 
\end{align}
where the upper and lower signs corresponds to either the first or the second set of parameters. This solution satisfies the following AKNS equation which is nothing else that the non-linear Schr\"odinger
\begin{equation}\label{akns2}
\Delta''-2|\Delta|^2\Delta+2 i m \cos\theta \Delta'+2m^2\Delta=0\,.
\end{equation}

The Shei complex kink and its properties are extensively in the literature, see for instance \cite{shei,BD}. In the context of a hidden supersymmetric structure associated with this solution, see  \cite{CDP}.

\item\textit{Coleman-Callan-Gross-Zee  (anti)kink}: This soliton appears as one of the most well-known solution of the $1+1$ dimensional GN$_2$ model \cite{dashen}. It can be obtained for $\theta=\pi/2$ from the above solution, which gives 
\begin{align}
\Sigma= \pm m \tanh \left(m x +\gamma \right), \quad M=0\, .
\end{align}
In this case the potential $\Delta$ is purely scalar and satisfies the following mKdV equation,
\begin{equation}\label{akns3}
\Delta''-2\Delta^3+2m^2\Delta=0\,.
\end{equation}

\item\textit{Dashen-Hassler-Neveu baryon}: Other interesting limit is when $\theta_1=\theta_2-\pi=\theta$, $\gamma_1=-\gamma_2=\tanh^{-1}\tan\theta/2 $, which leads to $\Sigma=0$ and
\begin{align}\label{dhnb}
\frac{M}{m}= 1-\sin \theta \left[ \tanh m \sin \theta \, x -\tanh \left(m\sin \theta x + \mu) \right) \right]
\end{align}
where $\mu=\tanh^{-1} \sin \theta$. In this case, the spectrum of the Hamiltonian becomes symmetric, i.e. the energies of the bound states have opposite sign $E_1=\cos \theta$, $E_2=-\cos \theta$. Consequently, considering $\Delta=M$ in (\ref{dhnb}) it satisfies the following mKdV equation,
\begin{equation}\label{akns4}
\Delta'''-6\Delta^2\Delta'+2 m^2 (\cos 2 \theta +2)\Delta'=0\,.
\end{equation} 
In the context of GN$_2$ this solution is known as the real Dashen-Hassler-Neveu baryon and represents a kink-antikink condensate\cite{dashen}.

\end{itemize}

The construction of the 2-twisted kink potentials by the Darboux transformation imposes the restriction $0 \leq \theta_1 < \theta_2 \leq \pi$.  This bound is in agreement with self-consistency conditions in the context of solutions of the NJL$_2$ model. The self-consistency  conditions relate the parameters $\theta_1$ and  $\theta_2$  with the fermion filling fraction of the
valence bound states. Once these conditions are set,  we can treat the 2-twisted kink as a solution of the gap equation in NJL$_2$. Hence. the results provide by the Darboux transformations are in concordance with the analysis given previously in the literature \cite{DT2,DT3, TN1, TN2}. Similar observations on the role of the Darboux transformation as a multi-soliton generating technique was done the context of solutions of the GN$_2$ and mKdV hierarchy  \cite{MikhailAdrian}.

\section{Symmetries of transparent Schr\"odinger Hamiltonians with matrix potential}\label{sec4}

Darboux transformation (\ref{L}), (\ref{intertwining}), and (\ref{htilde}) can be employed for construction of exactly solvable Schr\"odinger Hamiltonians with matrix potentials. These Hamiltonians appear in  a variety of physical situations. Let us mention coupled-channel scattering nuclear physics \cite{coupled}, the Jaynes-Cunning model in quantum optics \cite{jaynes}, or description of non-minimal coupling of neutral spin $1/2$ particles to the electromagnetic field \cite{Pronko},  see also the recent review \cite{Baye} and references therein. 

Taking the $H_0$ and $H$ from (\ref{h}) and (\ref{htilde}), respectively, we can define the following Schr\"odinger operators with matrix potentials
\begin{align}
 \mathcal{H}_0&=H_0^2=-\partial_{x}^2+\Sigma'_0 \,\sigma_3-M'_0\,\sigma_1+\Sigma_0^2+M_0^2 \, , \label{ache}
\\
\mathcal{H}&=H^2=-\partial_{x}^2+\Sigma'\, \sigma_3-M' \, \sigma_1+\Sigma^2+M^2 \label{achet}\, .
\end{align}
By construction, the spectra of  both $\mathcal{H}_0$ and $\mathcal{H}$ are non-negative and can differ in one or two discrete eigenvalues, in coherence with discussion provided in Section \ref{sec2}. The squared Hamiltonians are intertwined by the Darboux transformation $L$
\begin{equation}\label{intertwiningSchr}
 L\,\mathcal{H}_0=\mathcal{H}\,L\quad L^{\dagger}\, \mathcal{H}= \mathcal{H}_0\, L^{\dagger} \, .
\end{equation}
The Hamiltonians have also guaranteed the existence of integrals of motion, $[\mathcal{H},H]=0$ and $[\mathcal{H},H_0]=0$.
 
Let us notice that the intertwining relations of Schr\"odinger operators with matrix potentials and  their properties  have been systematically analyzed in the literature, see e.g. \cite{andrianov,samsonov,sokolov}. In this context, the situation considered here is more peculiar in the sense that Darboux transformation intertwines \textit{not only} the Schr\"odinger Hamiltonians $\mathcal{H}_0$ and $\mathcal{H}$, but \textit{also} their integrals of motion, the Dirac operators $H_0$ and $H$. 

\begin{figure}
\centering
\includegraphics[scale=.8]{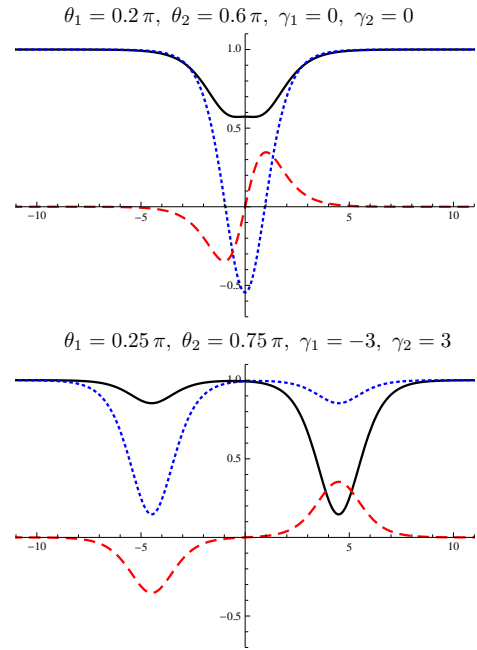}
\caption{Solid (black) line corresponds to $M^2+\Sigma^2+\Sigma'$, the dotted (blue) line corresponds to $M^2+\Sigma^2-\Sigma'$. The dashed (red) line corresponds to $M'$.}\label{fig2}
\end{figure}

We can use the results of the previous section taking the free particle system described by (\ref{free}) as the initial model, $\Sigma_0=0$ and $M_0=m$, 
\begin{equation}\label{matrixSchr}
 \mathcal{H}_0=H_0^2=-\partial_x^2+m^2,\quad H_0=i\sigma_2\partial_x+m\sigma_3\,.
\end{equation} 
Substituting to (\ref{achet}) the explicit form of $\Sigma$ and $M$ from (\ref{sigma}) and (\ref{eme}), we get a four-parameter family of exactly solvable \emph{matrix} Schr\"odinger Hamiltonians $\mathcal{H}$ with transparent potentials. These systems describe interesting physical situation:  for $|x| \rightarrow \infty$, the potential terms proportional to $\sigma_1$ and $\sigma_3$ vanish whereas $\Sigma^2+M^2$ is constant. Hence, $\mathcal{H}$ describes an asymptotically free particle. It undergoes a well localized, spin-sensitive, interaction that mixes its spin-up and spin-down components (or the two dynamical channels) whenever the term $M'\sigma_1$ is non-vanishing. 
For illustration of the considered family of potentials, see Figure~\ref{fig2}.

When compared to $H_0$, the Hamiltonian $\mathcal{H}_0$ possesses a richer set of integrals of motion. It has the following set of local symmetries
\begin{equation}\label{intSchr}
 \sigma_a,\quad p_0=-i\partial_x\quad\mbox{and}\quad p_a\equiv-i\sigma_a\partial_x,\quad a=1,2,3\,.
\end{equation}
\begin{eqnarray}\label{a1}
\,[\mathcal{H}_0,\sigma_a]=0,\quad [\mathcal{H}_0,p_a]=0\,. 
\end{eqnarray}
They close a Lie algebra where the structure coefficients are energy-dependent. The momentum operator $p_0$ plays the role of the central element. The non-vanishing commutation relations read 
\begin{align}\notag
&[\sigma_a,\sigma_b]=2i\epsilon_{abc}\sigma_c,
\quad 
[p_a,p_b]=2i\epsilon_{abc}\sigma_c (\mathcal{H}_0-m^2),& \\
& [\sigma_a,p_b]=2i\epsilon_{abc}p_c\, .&\label{a2}
\end{align}
In terms of the operators (\ref{intSchr}), the Dirac Hamiltonian $H_0$ can be written as the linear combination $H_0=-p_2+m\sigma_1$ and satisfies the following commutation relations
\begin{align}\label{hps}
 [H_0,p_a]&=[-p_2+m\sigma_1,p_a] \notag \\ 
 &=2i\epsilon_{2ac}\sigma_c(\mathcal{H}_0-m^2)-2im\epsilon_{a1c}p_c \, ,\\
 [H_0,\sigma_a]&=-2i\epsilon_{2ac}p_c+2im\epsilon_{1ac}\sigma_c\,.
\end{align}

Let us consider now the algebraic structure of the integrals of motion of the Hamiltonian $\mathcal{H}$. 
We can define the symmetry operators 
\begin{equation}\label{lpl}
\tilde{\sigma}_a=L\sigma_aL^{\dagger},\quad \tilde{p}_{\mu}=Lp_{\mu}L^{\dagger}\, ,\end{equation}
\begin{equation}
 [\tilde{\sigma_a},\mathcal{H}]=[\tilde{p}_{\mu},\mathcal{H}]=0\, .
\end{equation}
where $ a=1,2,3,$ $\mu=0,1,2,3.$
The algebraic structure closed by $\tilde{p}_a$ and $\tilde{\sigma}_a$ is more complicated when compared to the formulas (\ref{a2}). It is caused by the fact that the operators (\ref{intSchr}) do not commute with the Dirac Hamiltonian $H_0$. To illustrate the situation, let us consider  the commutator between $\tilde{\sigma}_a$ and $\tilde{\sigma}_b$. Keeping in mind the definition (\ref{lpl}), we can write 
\begin{align}\label{comut}
 &[\tilde{\sigma}_a,\tilde{\sigma}_b]=\\ \notag
 &2i\epsilon_{abc}\tilde{\sigma}_cP(H)
-(E_1+E_2)L\left(\sigma_a[H_0,\sigma_b]-\sigma_b[H_0,\sigma_a]\right)L^{\dagger}\,.
\end{align}
With the use of (\ref{hps}), we can rewrite the second term as a linear combination of the known integrals. This way, we get the following algebra
\begin{align}\label{lie}
& [\tilde{\sigma}_i,\tilde{\sigma}_j]=
2i\epsilon_{ijk}\tilde{\sigma}_kP(H)-2(E_1+E_2)
\left(
-\delta_{2i}\tilde{p}_j+\delta_{2j}\tilde{p}_i\right)\nonumber\\
&-2(E_1+E_2)\left(m(\delta_{1i}\tilde{\sigma}_{j}-\delta_{1j}\tilde{\sigma}_i)\right)\nonumber\\
&-4i(E_1+E_2)\left(\epsilon_{2ij}\tilde{p}_0-2im\epsilon_{1ij}\tilde{\sigma}_0\right)\,,\\
& \, [\tilde{p}_i,\tilde{p}_j]=
[\tilde{\sigma_i},\tilde{\sigma}_j](\mathcal{H}-m^2)\,,\\
&[\tilde{\sigma}_i,\tilde{p}_j]=
2i\epsilon_{ijk}P(H)\tilde{p}_k+4i(E_1+E_2)m\epsilon_{1ij}\tilde{p}_0\nonumber\\
&-2(E_1+E_2)
(-\delta_{2i}\tilde{\sigma}_j+\delta_{2j}\tilde{\sigma}_i)(\mathcal{H}-m^2)\nonumber\\&-2(E_1+E_2)\left(m(\delta_{1i}\tilde{p}_{j}-\delta_{1j}\tilde{p}_i)
+2i\epsilon_{2ij}(\mathcal{H}-m^2)\right).
\end{align}
We can see that when $E_1=-E_2$, the second term in the commutator cancels out and the operators close a Lie algebra which is a deformation of  (\ref{a2}) with  energy-dependent structure coefficients.

\section{Conclusion and further problems}

In this paper, distinct aspects of Darboux transformations and their applications into different models were considered.  A four-parameter family of reflectionless (transparent) potentials was derived by means of a Darboux transformation applied on the free particle Dirac Hamiltonian. Using this technique, static scattering solutions of the NJL$_2$ model can be found. Several comments are in order:

\begin{itemize}[itemindent=10pt,leftmargin=0pt]

\item The Darboux transformations (\ref{L}) can be utilized to generalize a hierarchy of transparent Hamiltonians. This can be done by applying a consecutive chain of such Darboux transformations. Let us suppose that $L_1$ intertwines the initial Hamiltonian $H_0$ with $H$.  The Hamiltonian $H$ can have one or two additional bound states when compared to $H_0$, in dependence on the explicit choice of the vectors $u_1$ and $u_2$, see (\ref{v}) and the discussion below. We can take $H_1\equiv H$ as the initial system now and define the new Darboux transformation $L_2$ and the new Hamiltonian $H_2$ that satisfies $L_2H_1=H_2L_2$. Thus, we also have $L_2L_1H_0=H_2L_{2}L_1$. The Hamiltonian $H_2$ can have up to four additional bound states when compared to $H_0$. In this way, we can create a whole hierarchy of transparent Hamiltonians with an arbitrary number of bound states. $H_1$ will appear as an intermediate Hamiltonian  between $H_0$ and $H_2$. The process here should be similar to the one known in non-
relativistic reflectionless operators \cite{Matveev}.

\item The 2-twisted kink potential analyzed here is based on the specific choice of the seed states in \eqref{b11}. However, the stationary equation $H_0f=E f$ has two independent solutions for any $E$. Therefore, other choices of $u_1$ and $u_2$, corresponding to the same energies $E_{1(2)}$ (as linear combinations for example), would be possible as well. Naturally, this would lead to different intertwining operators and to a new Hamiltonian $H$ as well. 

\item Having in mind the two points above, it will be interesting to analyze more generic situation, e.g. the implementation Darboux transformations in the context of time-dependent systems. It could generate the full, time dependent scattering solutions of the GN$_2$ and NJL$_2$ models, as in refs. \cite{DT,DT2,DT3}. In this spirit, it would be intriguing to consider the meaning of time-dependent objects, i.e.  breathers as pseudoscalar potentials, in the context of hidden symmetries.

\end{itemize}

Construction of exactly solvable Dirac equations via Darboux transformation (\ref{L}) proves to be fruitful in various areas of physics. Let us notice the recent results where solvable reflectionless models of twisted carbon nanotubes were constructed in this manner \cite{susytwisting} and discussed in the context of AKNS hierarchy \cite{AKNStwistedtubes}. In this context, the exactly solvable Hamiltonians (\ref{htilde}) can be interpreted as the effective Hamiltonian of Dirac fermions in graphene. The potential terms $M$ and $\Sigma$ then describe external magnetic field and an effective position-dependent mass, see e.g. \cite{VJKD} and references therein.  \\

As an aditional application of Darboux transformations, we showed that static transparent Schr\"odinger Hamiltonians can be derived as the square Dirac static transparent ones. 
The approach presented here resembles the framework of the supersymmetric quantum mechanics. There, the Schr\"odinger Hamiltonian is obtained as a square of the supercharge which is a first-order differential matrix operator with scalar potential $W$. When $W$ solves one of the equations of the mKdV hierarchy (and, hence, is transparent), then the diagonal elements $V_{\pm}$ of the potential $V=W^2+\sigma_3 W'$ of the Schr\"odinger Hamiltonian are also transparent and solve one of the equations of the KdV hierarchy. Let us notice that in the context of integrable systems, the relation $V_{\pm}=W^2\pm W'$ between the solution $W$ of mKdV hierarchy and the solutions $V_{\pm}$ of KdV hierarchy is known as Miura transformation.

The matrix Schr\"odinger Hamiltonians studied here are extension of this scheme. Here, we derive Schr\"odinger Hamiltonian with transparent potential as a square of Dirac operator with a pseudo-scalar potential that solves the AKNS hierarchy. The construction also provides integrals of motion like $\tilde{p}_0=L p_0 L^\dagger$ that can be identified with the Lax integral of motion of the matrix Schr\"odinger Hamiltonians. In this sense, these non-relativistic Hamiltonians with matrix potentials can be treated as integrable.

It is worth noticing that exact solvability of Schr\"odinger operators with matrix potential was discussed in the context of shape-invariance in  \cite{SIothers} or \cite{Nikitin}, where the classification of these systems was presented. The shape invariance of Dirac operators in the context of the Darboux transformation (\ref{L}) was discussed in \cite{CarvedFullerenes}. 
It would be interesting to analyze whether Schr\"odinger operators with the reflectionless matrix potentials can possess a wider set of (dynamical) symmetries. The works \cite{Negro} or \cite{Negro1} where dynamical symmetries are considered for transparent P\"oschl-Teller potential could serve as a starting point in this direction. 

\section*{Acknowledgments}
FC is partially supported by the Fondecyt grant 11121651 and by the Conicyt grant  79112034. FC  thanks the kind hospitality of the Nuclear Physics
Institute of the Academy of Sciences of the Czech Republic. VJ thanks The Centro de Estudios Cient\'{\i}ficos (CECs) in Valdivia, Chile, for hospitality. He is also  grateful to Javier Negro for discussions. The Centro de Estudios Cient\'{\i}ficos (CECs) is funded by the Chilean Government
through the Centers of Excellence Base Financing Program of Conicyt.



\begin{thebibliography}{999}
\bibitem{footnote1}{Here we use the units $\hbar =c=1$ and $\sigma _i$, $i=1,2,3$ are the sigma Pauli matrices.}
\bibitem{footnote2}{ A different construction of Darboux transformation in the context of AKNS systems was discussed in \cite{manies} where it was provided in terms of Grammian-type determinants of vector solutions of the Lax pairs.}
\bibitem{footnote3}{In those cases, the advantage of intertwining relations is lost as the Darboux transformation maps physical states into non-physical ones. Nevertheless, the creation of such a singularities using Darboux transformations can be circumvented considering, for instance, quantum $\protect \mathcal  {PT}$-symmetric systems \cite {ptsing}.}
\bibitem{footnote5}{Notice that the spinor part of $v_1$ and $v_2$ depends on $\theta _2$ and $\theta _1$, respectively. Dependence on the parameters $\theta _1$ and $\theta _2$, respectively, is hidden in the determinant of the matrix $U$.}
\end{thebibliography}
\end{document}